\documentclass[12pt]{iopart}
\usepackage{iopams}  
\usepackage{color}
\usepackage{tabularx}
\usepackage{epsfig}
\usepackage{graphicx}

\begin{document}

\title{Hierarchical Regular Small-World Networks}

\author{Stefan Boettcher and Bruno Gon{\c c}alves}
\address{Emory University, Dept. of Physics, Atlanta, GA 30322, USA}

\author{Hasan Guclu }
\address{Center for Nonlinear Studies, Los Alamos National
Laboratory, MS-B258, Los Alamos, NM 87545, USA}

\begin{abstract}
Two new networks are introduced that resemble small-world
properties. These networks are recursively constructed but retain a
fixed, regular degree. They possess a unique one-dimensional lattice
backbone overlayed by a hierarchical sequence of long-distance links,
mixing real-space and small-world features. Both networks, one
3-regular and the other 4-regular, lead to distinct behaviors, as
revealed by renormalization group studies.  The 3-regular network is
planar, has a diameter growing as $\sqrt{N}$ with system size $N$, and
leads to super-diffusion with an exact, anomalous exponent
$d_{w}=1.306\ldots$, but possesses only a trivial fixed point
$T_{c}=0$ for the Ising ferromagnet. In turn, the 4-regular network is
non-planar, has a diameter growing as $\sim2^{\sqrt{\log_{2}N^{2}}}$,
exhibits {}``ballistic'' diffusion ($d_{w}=1$), and a non-trivial
ferromagnetic transition, $T_{c}>0$. It suggest that the 3-regular
network is still quite {}``geometric'', while the 4-regular network
qualifies as a true small world with mean-field properties. As an
engineering application we discuss synchronization of processors on
these networks.
\end{abstract}
\pacs{
89.75.-k, 
64.60.ae, 
64.60.aq, 
05.50.+q 
}
\maketitle
The description of the {}``6-degrees-of-separation'' phenomenon in
terms of small-world (SW) networks by Watts and Strogatz \cite{Watts98} has
captured the imagination of many researchers, and was particularly
timely as we suddenly found ourselves in a networked world
\cite{Boccaletti06,Barabsi03,Newman01}.  Such a rich environment
requires a diverse set of tools and models for their
understanding. Statistical physics, with its notion of universality,
provides powerful methods for the classification of complex systems,
like the renormalization group (RG)
\cite{Goldenfeld,Stanley99,Plischke94,Newman99}. 

\begin{figure}

\includegraphics[bb=10bp 50bp 850bp 613bp,clip,width=210pt,height=160pt]{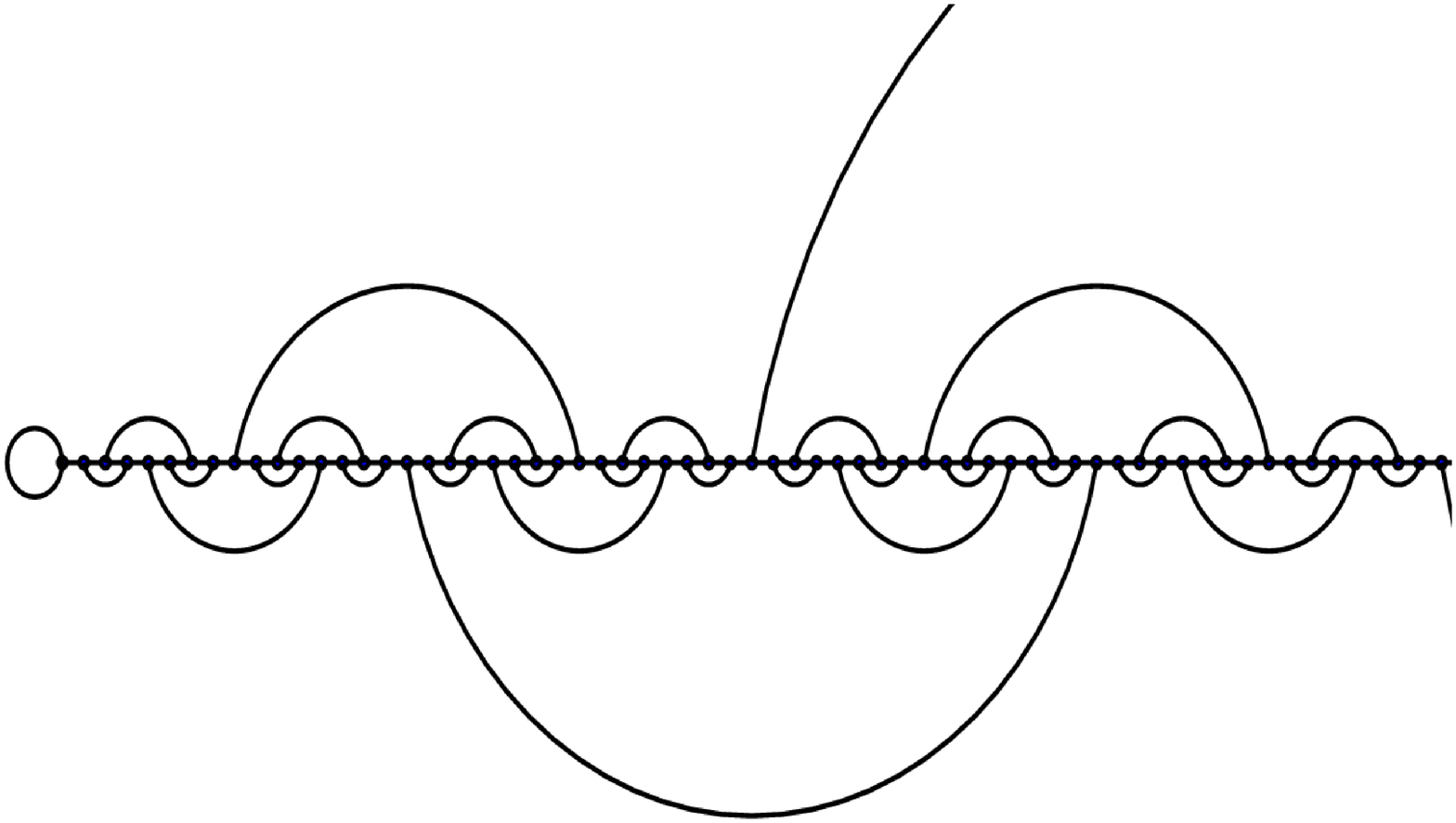}
\hfill
\includegraphics[bb=25bp 14bp 490bp 613bp,clip,width=150pt,height=220pt,angle=90]{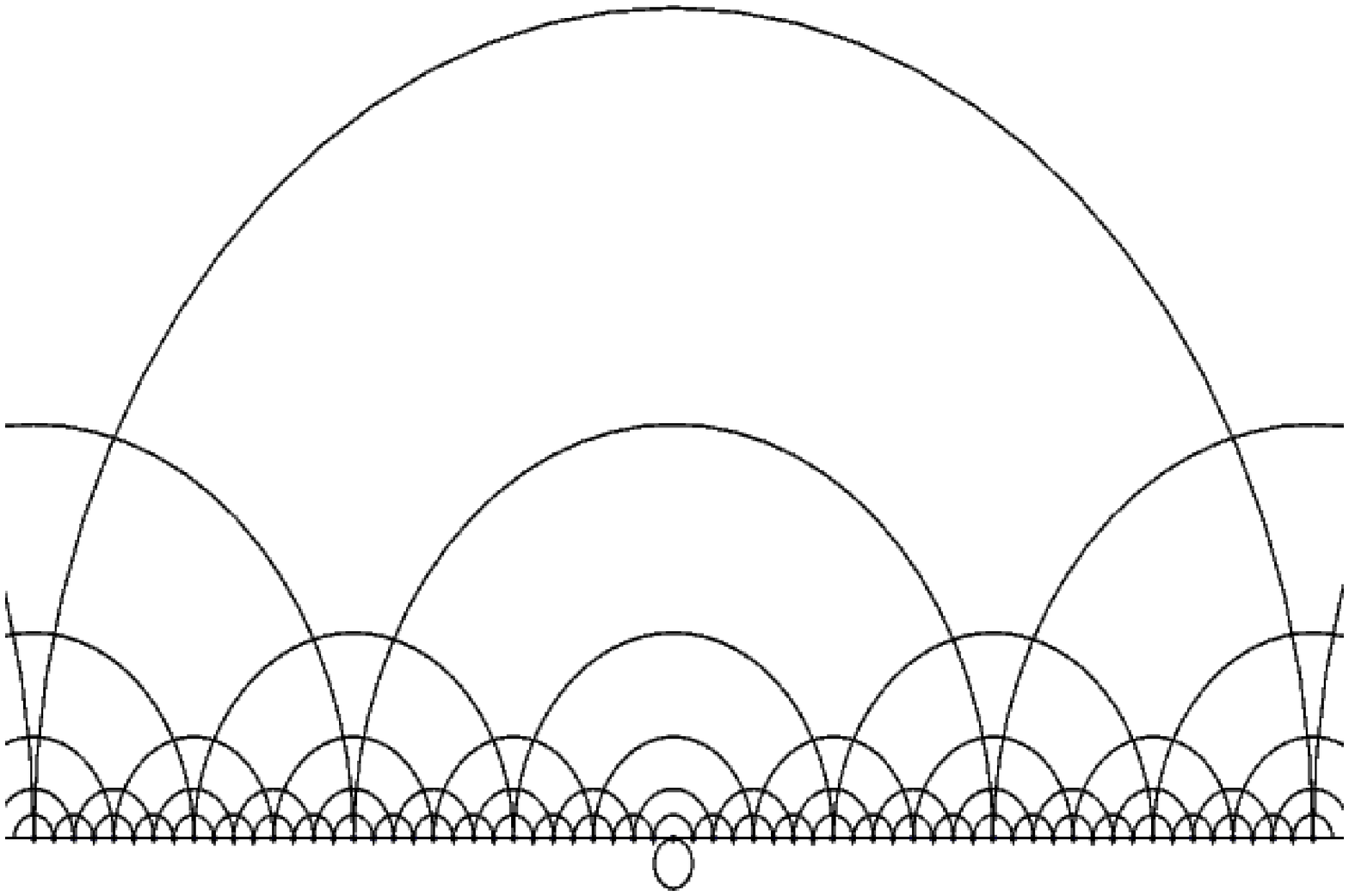}

\caption{\label{fig:3hanoi}
Display of the 3-regular network HN3 (left) and 4-regular network HN4
(right). HN3 is planar but HN4 is not.}
\end{figure}

Here, we introduce and study a set of graphs which reproduce
the behavior of SW networks without the usual disorder inherent in
natural networks.  Instead, they attain these properties in a
recursive, hierarchical manner that is conducive for RG. The
motivation is comparable to regular scale-free networks proposed in
Refs.~\cite{Barabasi01,Andrade05} or the Migdal-Kadanoff RG
\cite{Migdal76,Kadanoff76,Hinczewski06}. The benefit of these features
is two-fold: For one, we expect that many SW phenomena can be studied
analytically on these networks, and that they will prove as useful as,
say, Migdal-Kadanoff RG has been for physical systems in low
dimensions. Furthermore, possessing such well-understood and
\emph{regular} networks is a tremendous advantage for engineering
applications, as it is difficult to manufacture realizations of
\emph{random} networks reliably when we can ascertain their behavior
only in the ensemble average. Here, we introduce these networks by
discussing their geometry and physical processes on them, such as
diffusion, phase transitions, and synchronization.

Each network possesses a geometric backbone, a one-dimensional
line of $N=2^k$ sites, either open or closed into a ring. To obtain a
SW hierarchy, we  parameterize
\begin{eqnarray}
n & = & 2^{i}\left(2j+1\right)
\label{eq:numbering}
\end{eqnarray}
\emph{uniquely} for any integer $n\not=0$, where
$j=0,\pm1,\pm2,\ldots$ labels consecutive sites within each level
$i\geq0$ of the hierarchy. For instance, $i=0$ refers to all odd
integers, $i=1$ to all integers once divisible by 2 (i.e.,
$\pm2,\pm6,\pm10,\dots$), and so on. In these networks, both depicted
in Fig.~\ref{fig:3hanoi}, in addition to its nearest neighbor in the
backbone, each site is also connected with (one or both) of its
neighbors \emph{within} the hierarchy.  For example, we obtain a
hierarchical 3-regular network HN3 by connecting first neighbors in
the 1D-backbone, then 1 to 3, 5 to 7, 9 to 11, etc, for $i=0$, next 2
to 6, 10 to 14, etc, for $i=1$, and 4 to 12, 20 to 28, etc, for $i=2$,
and so on. The 4-regular network HN4 is obtained in the same manner,
but connecting to \emph{both} neighbors in the hierarchy. For HN4 it
is clearly preferable to extend the line to $-\infty<n<\infty$ and
also connect -1 to 1, -2 to 2, etc, as well as all negative integers
in the above pattern. These networks resemble models of
ultra-diffusion \cite{Huberman85,Ogielski85}, but with inhibiting
barriers replaced by short-cuts here.

\begin{figure}

\hfill
\includegraphics[scale=0.25,angle=90]{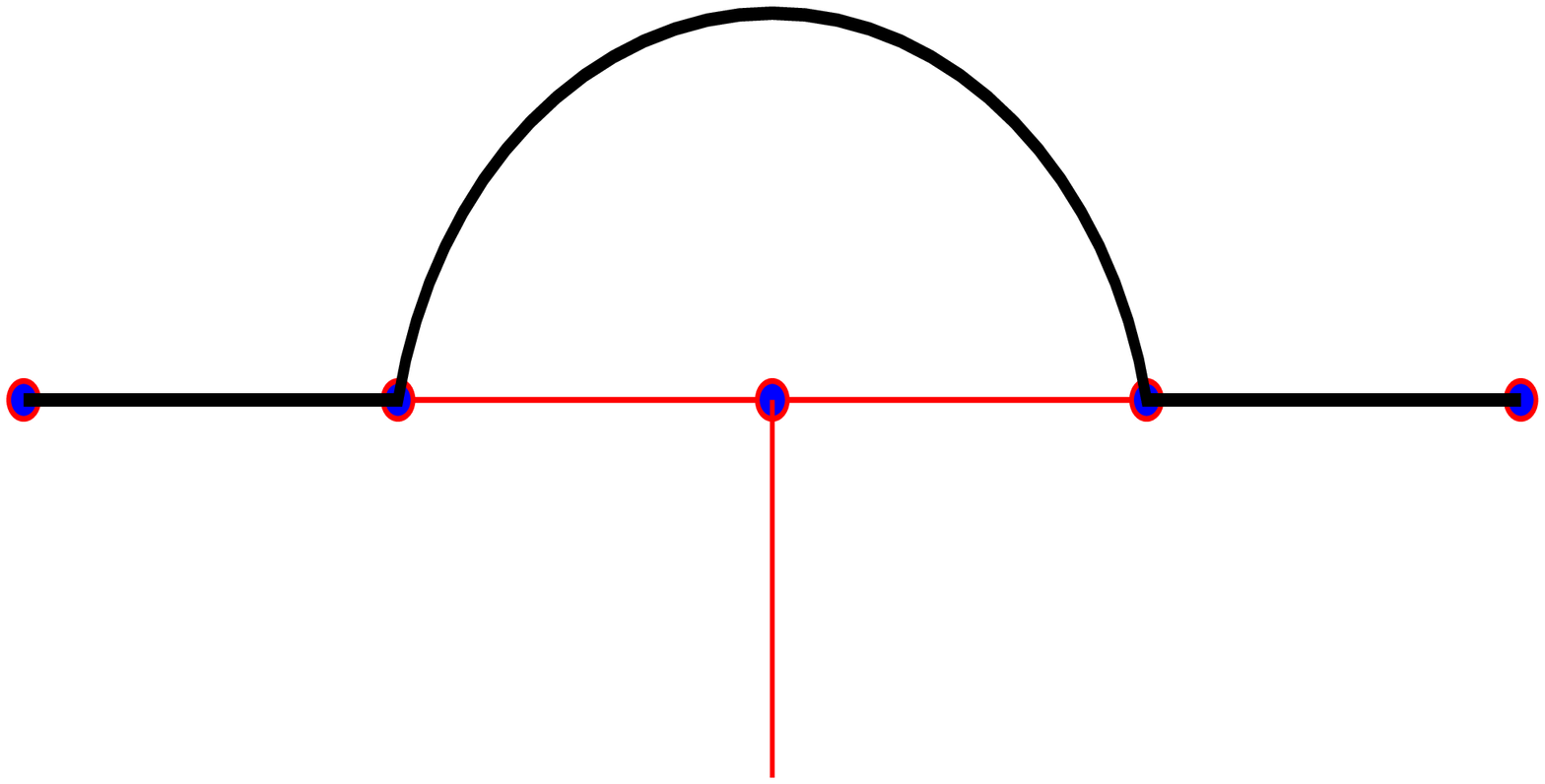}
\includegraphics[scale=0.25,angle=90]{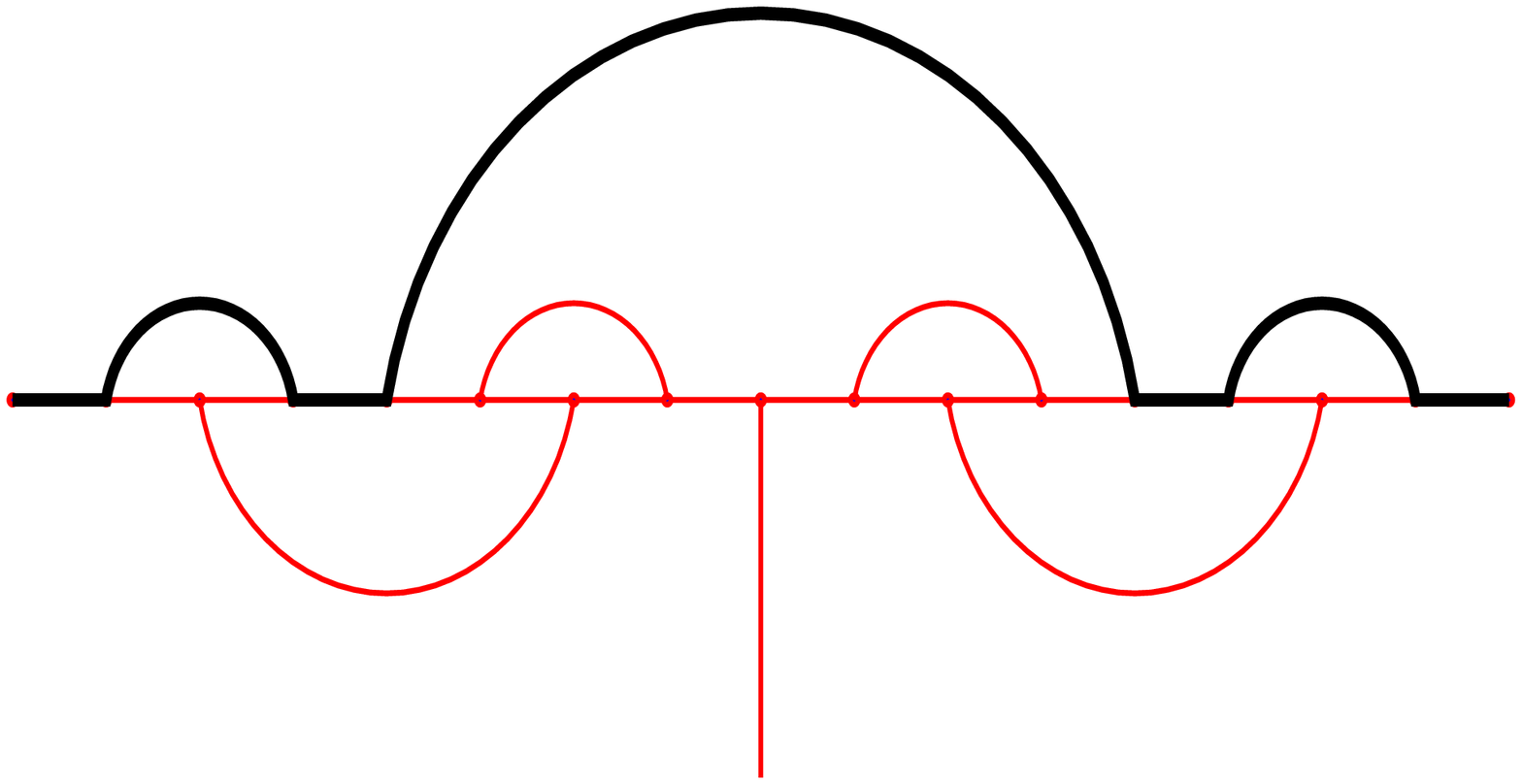}
\includegraphics[scale=0.25,angle=90]{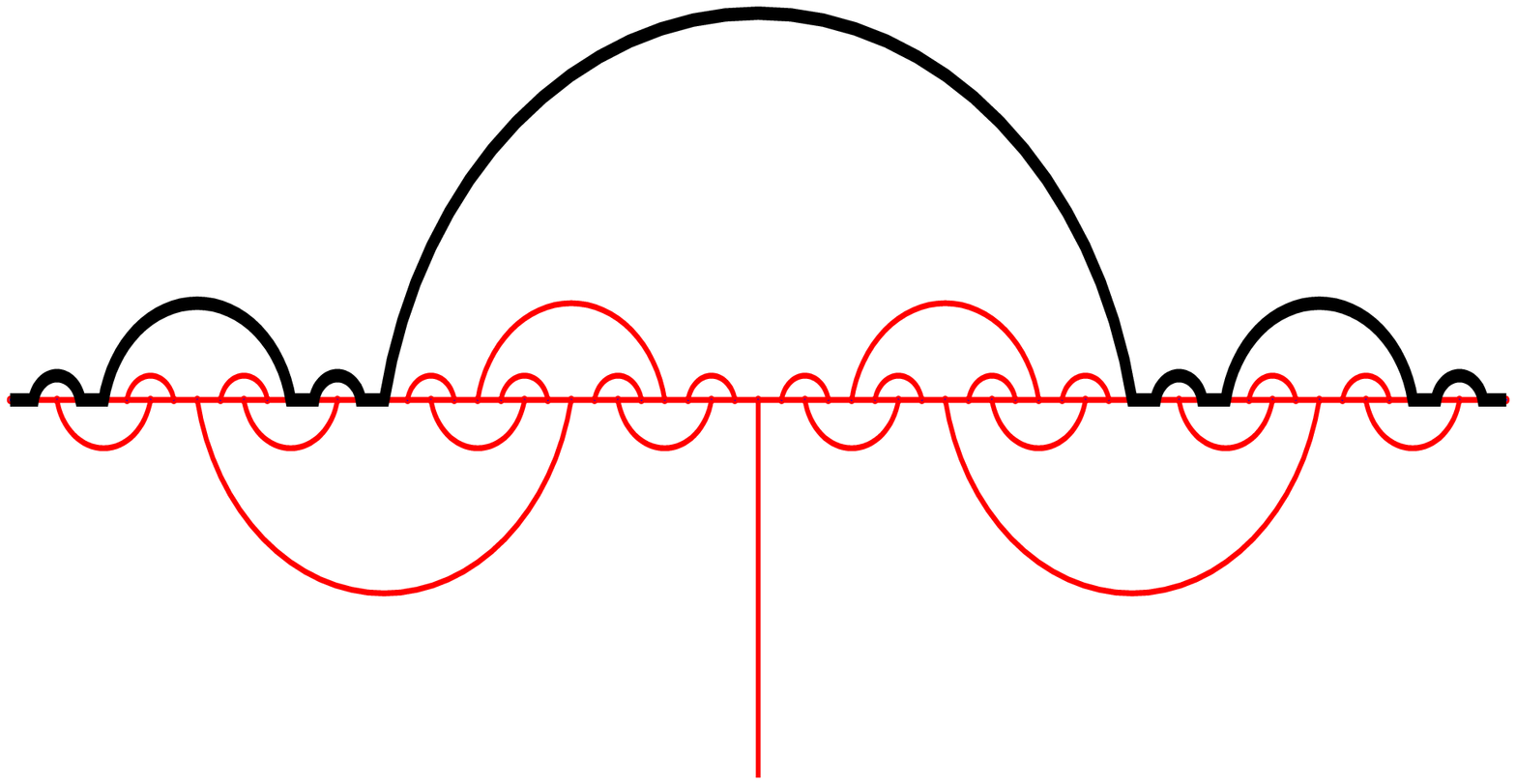}
\caption{\label{fig:Koch3hanoi} 
(Color Online)
Sequence of shortest end-to-end paths (=diameter, thick lines)
for HN3 of size $N=2^{k}$, $k=2,4,8$.  Whenever the system size $N$
increases by a factor of 4, the diameter $d$ increases by a factor of
$\sim2$, leading to Eq.~(\ref{eq:3dia}). }
\end{figure}

It is simple to determine geometric properties. For instance, both
networks have a clustering coefficient \cite{Boccaletti06} of $1/4$.
Next, we consider the diameter $d$, the longest of the shortest paths
between any two sites, here the end-to-end distance. Using system
sizes $N_k=2^k$, $k=2,4,6,\ldots$, for HN3, the diameter-path looks
like a Koch curve, see Fig.~\ref{fig:Koch3hanoi}. The length $d_{k}$
of each marked path is given by $d_{k+2}=2d_{k}+1$ for
$N_{k+2}=4N_{k},$ hence
\begin{equation}
d\sim\sqrt{N}.
\label{eq:3dia}
\end{equation}
This property is reminiscent of a square-lattice of $N$ sites, whose
diameter (=diagonal) is also $\sim\sqrt{N}$. HN3 is thus far from true
SW behavior where $d\sim\ln N$.

The geometry of HN4 is more subtle. We consider again the shortest
path between the origin $n=0$ and the end $n=N=2^{k}$. Due to
degeneracies at each level, one has to probe many levels in the
hierarchy to discern a pattern. In fact, any pattern evolves for an
increasing number of levels before it gets taken over by a new one,
with two patterns creating degeneracies at the crossover. We find that
the paths here do \emph{not} search out the longest possible jump, as
in Fig.~\ref{fig:Koch3hanoi}.  Instead, the paths reach quickly to
some intermediate level and follow \emph{consecutive} jumps at that
level before trailing off in the end. This is a key distinguishing
feature between HN3 and HN4: Once a level is reached in HN4, the
entire network can be traversed at that level, while in HN3 one
\emph{must} switch to lower levels to progress, see
Fig.~\ref{fig:3hanoi}. 

We derive a recursion equation~\cite{SWlong} with the solution 
\begin{eqnarray}
d & \sim\frac{1}{2} & \sqrt{\log_{2}N^{2}}\,\,2^{\sqrt{\log_{2}N^{2}}}
\qquad\left(N\to\infty\right)
\label{eq:dkasymp}
\end{eqnarray}
for the diameter of HN4. Expecting the diameter of a small world to
scale as $d\sim\log N$, we rewrite Eq.~(\ref{eq:dkasymp}):
\begin{eqnarray}
d_{k} & \sim & \left(\log_{2}N\right)^{\alpha}\quad{\rm with}
\quad\alpha\sim\frac{\sqrt{\log_{2}N^{2}}}{\log_{2}\log_{2}N^{2}}+\frac{1}{2}.
\label{eq:alpha}
\end{eqnarray}
Technically, $\alpha$ diverges with $N$ and the diameter
grows faster than any power of $\log_{2}N$ {[}but less than any
power of $N$, unlike Eq.~(\ref{eq:3dia})]. In reality, though, $\alpha$
varies only very slowly with $N$, ranging merely from
$\alpha\approx1.44$ to $\approx1.84$
over \emph{nine} orders of magnitude, $N=10-10^{10}$.

As a demonstration of the rich dynamics facilitated by these
networks, we have modeled diffusion on HN3 and HN4.
Starting a random walks at $n=0$, we focus here only on the mean
displacement with time,
\begin{equation}
\left\langle \left|n\right|\right\rangle \sim t^{1/d_{w}}.
\label{eq:MSD}
\end{equation}
All walks are controlled by the probability $p$ of a walker to step
off the lattice into the direction of a long-range jump. In
particular, the walker always jumps either to the left or right
neighbor with probability $\left(1-p\right)/2$, but makes a long-range
jump with probability $p$ on HN3, or $p/2$ to either the left or right
on HN4. In both cases, a simple 1D nearest-neighbor walk results for
$p=0$ with $d_{w}=2$ for ordinary diffusion.  For any probability
$p>0$, long-range links will dominate the asymptotic behavior, and
the leading scaling behavior becomes independent~of~$p$.

\begin{figure}[b!]

\hfill
\includegraphics[clip,scale=0.5]{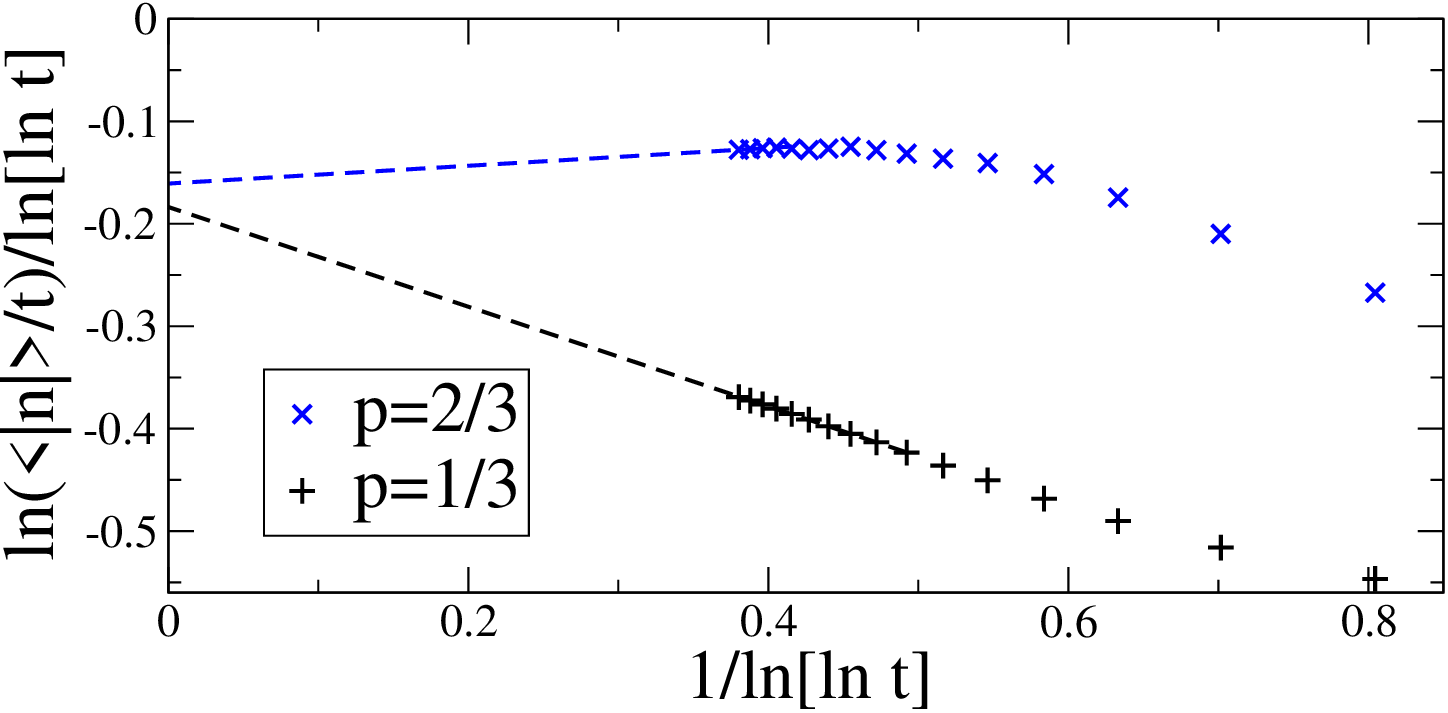}

\caption{\label{fig:RW4MSD}
Rescaled plot of the mean distance $\left\langle
\left|n\right|\right\rangle $ in HN4 for walks up to $t=10^{6}$. We
demonstrate that $d_{w}=1$ but with log-corrections by rewriting
Eq.~(\ref{eq:MSD}) as $\left\langle \left|n\right|\right\rangle /t\sim
V\left[\ln t\right]^{\beta}$.  Then we obtain $\ln\left(\left\langle
\left|n\right|\right\rangle /t\right)/\ln\left[\ln
  t\right]\sim\beta+\ln V/\ln\left[\ln t\right]$ and linearly
extrapolate (dashed lines) $1/\ln\left[\ln t\right]\to0$, estimating
$\beta\approx-0.18$ at the intercept, independent of $p$. An effective
{}``velocity'' $V$ could be extracted from the slope. For any value
besides $d_{w}=1$, these extrapolations would \emph{not} converge.}
\end{figure}

Adapting the RG for random walks in Refs.~\cite{Kahng89,Redner01}, we
find \emph{exact} results for HN3. The local analysis\cite{SWN} of the
physical fixed point is \emph{singular}, with a boundary layer instead
of a Taylor expansion, yielding an anomalous exponent of
$d_{w}=2-\log_2\left(\phi\right)=1.3057581\ldots$,
containing the (irrational) ``golden section''
$\phi=\left(1+\sqrt{5}\right)/2$. This is a remarkable exponent also
because it is a rare example of a simple walk with super-diffusive
($1<d_{w}<2$) behavior without Levy flights
\cite{Havlin87,Metzler04,Fischer07}, and it would be consistent with
experiments leading to super-diffusion \cite{Solomon93}. We have not
been able to extend this RG calculation to obtain analytic results for
HN4 yet, although the high degree of symmetry inherent in these
networks (and the simple result obtained) suggest the possibility. For
HN4, an annealed approximation and simulations, evolving some
$2\times10^{7}$ walks for $10^{6}$ time steps each, suggest a value of
$d_{w}=1$, see {}Fig.~\ref{fig:RW4MSD}. Hence, a walk on HN4 proceeds
effectively ballistic, but hardly with linear motion: widely
fluctuating jumps conspire just so that a single walker extends
outward with an on-average constant velocity in \emph{both}
directions, yet the walk remains recurrent. Clearly, it is easier to
traverse HN4 than HN3 because of the above-stated fact that on HN4 a
walker can progress \emph{repeatedly} within a hierarchical level.

We have also studied Ising spin models on HN3 and HN4, with RG and
with Monte Carlo simulations. First, we consider the RG for the Ising
model on HN3. In this case, all steps can be done exactly but the
result turns out to be trivial (for uniform bonds) in the sense that
there are no finite-temperature fixed points of the RG flow. Yet, the
calculation is instructive, highlighting the large number of
statistical models that can be accessed through the hierarchical
nature of the process, and it is \emph{almost} identical in outcome to
the treatment below for HN4. That small difference is just enough to
provide HN4 with a non-trivial $T_{c}>0$, which we confirm
numerically.

\begin{figure}

\hfill
\includegraphics[bb=0bp 560bp 375bp 720bp,clip,scale=0.60]{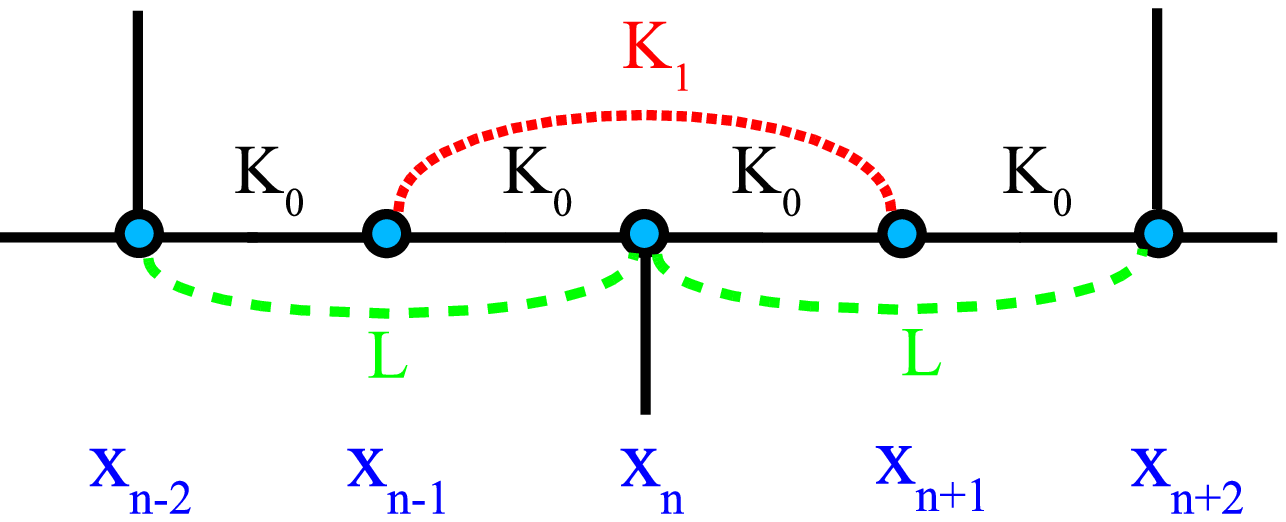}
\includegraphics[bb=60bp 560bp 300bp 720bp,clip,scale=0.60]{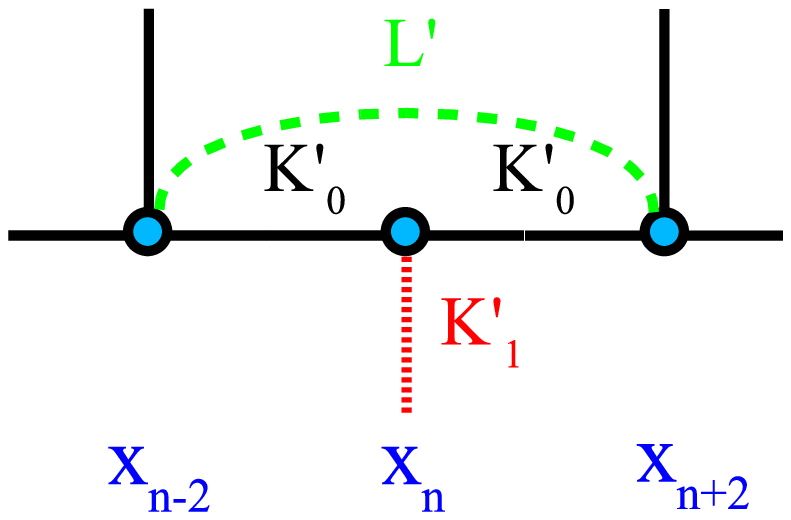}

\caption{\label{fig:RG3}
Depiction of (exact) RG step for the Ising model on HN3. Tracing out
odd-labeled variables $x_{n\pm1}$ for all $n=2(2j+1)$, in the left
plot leads to the renormalized couplings $(L',K'_{0},K'_{1})$ on the
right in terms of the old couplings $(L,K_{0},K_{1})$. Unlabeled bonds
correspond to $K_{i\geq2}$. HN3 does not contain couplings of type
$(L,L')$, but they become relevant during the RG process. Random walks
on HN3 lead to a topologically equivalent, but more involved RG step
\cite{SWN}.}
\end{figure}

The RG consists of recursively tracing out odd-relabeled spins
$x_{n\pm1}$, see Fig.~\ref{fig:RG3}. The $x_{n\pm1}$ are connected to
their even-labeled nearest neighbors on the lattice backbone by a
coupling $K_{0}$.  At any level, each $x_{n\pm1}$ is also connected to
one other such spin $x_{n\mp1}$ across an even-labeled spin $x_{n}$
with $n=2\left(2j+1\right)$ in Eq. (\ref{eq:numbering}) that is
exactly \emph{once} divisible by 2. Let us call that coupling $K_{1}$,
all other couplings are $K_{i>1}$.  During the RG process, a new
coupling $L$ (dashed line in Fig. \ref{fig:RG3}) between
next-nearest even-labeled neighbors emerges. Putting all higher level
terms into $\mathcal{R}$, we can section the Hamiltonian
\begin{eqnarray}
-\beta\mathcal{H} & = & \sum_{\left\{ n=2(2j+1)\right\} }
\left(-\beta\mathcal{H}_{n}\right)+\mathcal{R}\left(K_{2},K_{3},\ldots\right),
\label{eq:3Hamiltonian}
\end{eqnarray}
where each sectional Hamiltonian  is given by
\begin{equation}
-\beta\mathcal{H}_{n}=\sum_{m=n-2}^{n+1}K_{0}x_{m}x_{m+1}
+K_{1}x_{n-1}x_{n+1}+L\left(x_{n-2}x_{n}+x_{n}x_{n+2}\right)\nonumber
\label{eq:3HSection}
\end{equation}
with $\left(K_{0},K_{1},L\right)$ as unrenormalized couplings
and we neglected an overall energy scale. After tracing out the
odd-labeled spins in each $\exp\left[-\beta\mathcal{H}_{n}\right]$, we
identify the renormalized couplings (neglecting $I'$):
\begin{eqnarray}
K'_{0} &   =   & L + \frac{1}{2}\ln\cosh\left(2K_{0}\right) + \frac{1}{4}\ln\left[1
 + \tanh\left(K_{1}\right)\tanh^{2}\left(2K_{0}\right)\right],\nonumber \\
L' &   =   & \frac{1}{4}\ln\left[1
 + \tanh\left(K_{1}\right)\tanh^{2}\left(2K_{0}\right)\right],
\label{eq:3RG}
\end{eqnarray}
and $K'_{i}=K_{i+1}$ f.~a. $i\geq1$. The high-$T$ solution
$K_{0}^{*}=L^{*}=0$ is a trivial fixed point of
Eq.~(\ref{eq:3RG}). Excluding that and eliminating $L^{*}$ yields
$1=\tanh\left(K_{1}\right)\tanh\left(2K_{0}^{*}\right)$, which has
only the $T_{c}=0$ solution $K_{0}^{*}=\infty$ (where also
$K_{1}=J_{1}/T\to\infty$).  Note, however, that the RG
recursions~(\ref{eq:3RG}) have a remarkable property due to the
hierarchical structure of the network: The next-level coupling $K_{1}$
appears as a \emph{free parameter} and acts as ``source term'' that
could be chosen to represent physically interesting situations, e.~g.
disorder or distance-dependence. For instance, with $K_{i}$ as an
increase function of distance $r_{i}=2^{i+1}$, a non-trivial fixed
point could be created.

In contrast, HN4 provides a non-trivial solution for the Ising model
even for uniform bonds, as expected for a mean-field system. Again, an
exact result for HN4 is elusive, although in light of the inherent
symmetries such a solution appears possible. Instead, we proceed
to a Niemeijer-van Leeuwen cumulant expansion \cite{Plischke94} and
compare with our numerical simulations. The Hamiltonian
 indeed has an elegant hierarchical form separating
the lattice backbone and long-range couplings:
\begin{eqnarray}
-\beta\mathcal{H}=\sum_{n=1}^{2^{k}}K_{0}x_{n-1}x_{n}+\sum_{i=1}^{k}
\sum_{j=1}^{2^{k-i}}K_{i}x_{2^{i-1}\left(2j-1\right)}x_{2^{i-1}\left(2j+1\right)}.
\label{eq:4Hamiltonian}
\end{eqnarray}
For the RG, we set 
$-\beta\mathcal{H}=-\beta\mathcal{H}_{0}-\beta\mathcal{V}+\mathcal{R}$
with
\begin{eqnarray}
-\beta\mathcal{H}_{0} & = & \sum_{j=1}^{2^{k-1}}K_{0}x_{2j-1}\left(x_{2j-2}+x_{2j}\right)
+\sum_{j=1}^{2^{k-1}}Lx_{2j-2}x_{2j},\nonumber \\
-\beta\mathcal{V}& = &\sum_{j=1}^{2^{k-1}}K_{1}x_{2j-1}x_{2j+1},
\label{eq:4Hdecomp}
\end{eqnarray}
adding new couplings $L$ that emerge during RG, as in
Fig.~\ref{fig:RG3}.  Tracing out odd spins and relabeling all
remaining even spin variables $x_{n}\to x'_{n/2}$, the cumulant
expansion applied to Eq.~(\ref{eq:4Hdecomp}) yields a new Hamiltonian
$-\beta\mathcal{H}'$, formally \emph{identical} to
Eq.~(\ref{eq:4Hamiltonian}), with the rescaled couplings
\begin{eqnarray}
K'_{0} & = & L+\frac{1}{2}\ln\cosh\left(2K_{0}\right)
+\frac{K_{1}}{2}\tanh^{2}\left(2K_{0}\right),\\
K'_{1} & = & K_{2}+\frac{K_{1}}{4}\tanh^{2}\left(2K_{0}\right),\quad
L'=\frac{K_{1}}{4}\tanh^{2}\left(2K_{0}\right).\nonumber 
\label{eq:RG4}
\end{eqnarray}
and $K'_{i}=K_{i+1}$ for $(i\geq2)$.  These are the same relation one
would obtain for the 1D-Ising model with \emph{nnn} couplings, if
$K_{1}\equiv L$ and $K_{i}=0$ for $i\geq2$. In that case, one would
find -- correctly -- that there are \emph{no} non-trivial fixed
points. The $K_{2}-$term, which appears as an arbitrary
source again at every RG step, if chosen appropriately, provides the
sole ingredient for a non-trivial outcome.  But unlike for HN3, here
already uniform ($i$-independent) $K_{i}$ obtain $T_{c}>0$. Holding
the source terms fixed, $K_{i\geq2}\equiv1$, we find a single
nontrivial fixed point at $K_{0}^{*}\approx0.2781$,
$K_{1}^{*}\approx1.0681$, $L^{*}\approx0.0681$. An analysis of the RG
flow \cite{Plischke94} in Eqs.~(\ref{eq:RG4}), starting with identical
$K_{i}\equiv\beta J$ f.~a.~$i$, yields
$T_{c}\approx2.2545J$. Simulations on HN4 with uniform bonds for
increasing system sizes $N=2^{k}$ accumulate to $T_{c}=2.1(1)J$.

\begin{figure}

\hfill
\includegraphics[clip,scale=0.5]{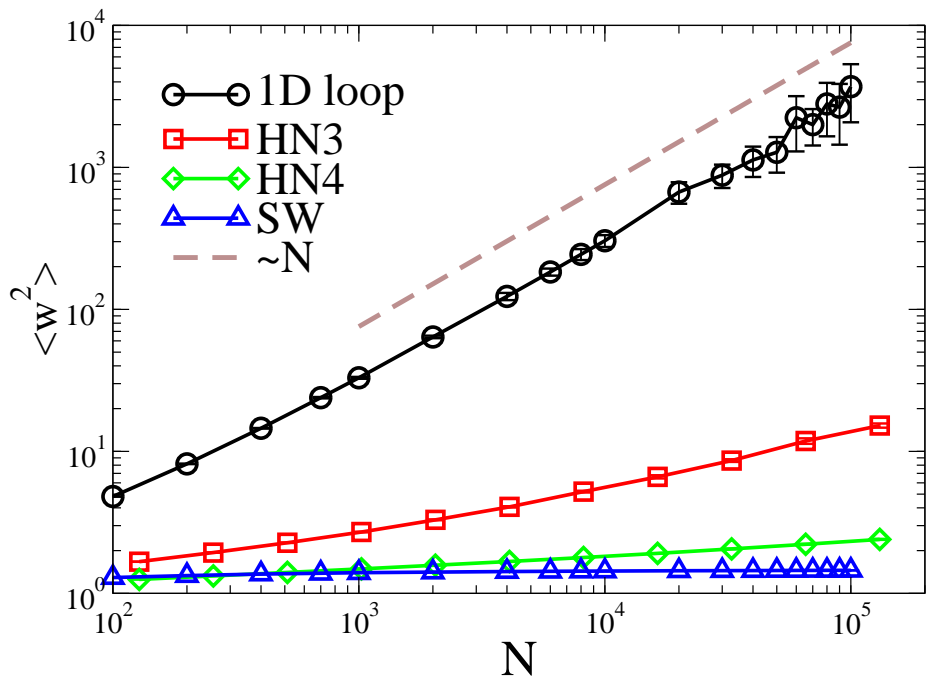}
\vspace{-0.2cm}
\caption{\label{fig:SyncroSW} 
Width $\left<w^{2}\right>$ as a function of $N$.  The 1D-loop without
long-range connections diverges most strongly, linear in $N$, while
the random one-per-node SW connections keep the width finite. The
width diverges with a weak power of $N$ for HN3, but merely
logarithmically for HN4. }
\vspace{-0.5cm}
\end{figure}

Finally, we demonstrate the usefulness of having a regular (i.e.,
non-random) network at hand with fixed, \emph{predictable}
properties. Synchronization is a fundamental problem in natural and
artificially coupled multicomponent systems \cite{Strogatz01}. Since
the introduction of SW networks \cite{Watts98}, it has been
established that such networks can facilitate autonomous
synchronization \cite{PhysRevLett.89.054101,Korniss03}. In a
particular synchronization problem the nodes are assumed to be task
processing units, such as computers or manufacturing devices. Let
$h_{i}(t)$ be the total task completed by node $i$ at time $t$ and the
set $\{ h_{i}(t)\}_{i=1}^{N}$ constitutes the task-completion
(synchronization) landscape, where $N$ is the number of nodes.  In
this model the nodes whose tasks are smaller than those of their
neighbors are incremented by an exponentially distributed random
amount, i.e., the node $i$ is incremented, if $h_{i}(t)\leq\min_{j\in
  S_{i}}\{ h_{j}(t)\}$, where $S_i$ is the set of nodes connected to
node $i$; otherwise, it idles.  In its simplest form the evolution
equation is $h_{i}(t+1)=h_{i}(t)+\eta_{i}(t)\prod_{j\in
  S_{i}}\Theta \left(h_{j}(t)-h_{i}(t)\right)$, with
\emph{iid} random variables of unit mean, $\eta_{i}(t)$, $\delta$-correlated in
space and time, and $\Theta$ as the Heaviside step function.

The average steady-state spread or width of the synchronization
landscape (degree of de-synchronization) can be written as $w^{2} =
(1/N)\sum_{i=1}^{N}(h_{i}-\bar{h})^{2}$ \cite{Korniss03}.  In low
dimensional regular lattices the synchronization landscape belongs to
the Kardar-Parisi-Zhang \cite{KPZ} universality class, a rough
desynchronized state dominated by large-amplitude long-wavelength
fluctuations, where width diverges with $N$. On the contrary, the
width becomes finite \cite{Korniss03} on a SW model in which each node
is connected to nearest neighbors and one random neighbor. In
Fig.~\ref{fig:SyncroSW}, we show the width as a function of $N$ for a
1D loop and SW, as well as HN3 and HN4.  The width for HN4 behaves
very similar to SW, with at most a logarithmic divergence in $N$,
while it diverges with power-law for HN3, but weaker than for a
1D-loop. HN4, thus, provides very similar properties to SW with the
benefit of a regular and reproducible structure that is easy to
manufacture, and that is potentially analytically tractable.

In conclusion, we introduced a new set of hierarchical networks with
regular, small world properties and demonstrated their usefulness for
theory and engineering applications with  a few examples.
Aside from the countless number of statistical models that can be
explored with RG on these networks, they also provide a systematic way
to interpolate off a purely geometric lattice into the SW
domain, possibly all the way into the mean-field regime (for
HN4). Even though at this point complete solutions on HN4 elude the
authors, even the leading approximation provides significant insight.

B.G. was supported by the NSF through grant \#0312510 and
H.G. was supported by the U.S. DOE through DEAC52-06NA25396.

\section*{References}
\bibliographystyle{unsrt}
\bibliography{/Users/stb/Boettcher}

\end{document}